# Scrape-off layer ion temperature measurements at the divertor target in MAST by retarding field energy analyser

S. Elmore[a,b,*], S. Y. Allan[a], A. Kirk[a], A. J. Thornton[a], J. R. Harrison[a], P. Tamain[c], M. Kočan[d] , J. W. Bradley[b] and the MAST Team[a]

[a]*EURATOM/CCFE Fusion Association, Culham Science Centre, Abingdon, Oxon, OX14 3DB, UK.*

[b]*University of Liverpool, Brownlow Hill, Liverpool, L69 3GJ, UK.*

[c]*Association Euratom-CEA, CEA/DSM/IRFM, CEA-Cadarache, F-13108 St Paul-lez-Durance Cedex, France.*

[d]*Max-Planck Institut für Plasmaphysik, EURATOM Association, Garching, Germany.*

## Abstract

Knowledge of the ion temperature ($T_i$) is of key importance for determining heat fluxes to the divertor and plasma facing components, however data regarding this is limited compared to electron temperature ($T_e$) data. $T_i$ measurements at the divertor target, between edge-localised modes (inter-ELM) H-mode, have been made using a novel retarding field energy analyser (RFEA). Unlike previous L-mode measurements where $T_i = T_e$, results from a range of H-mode discharges have shown that at the target $T_i/T_e = 1.5 - 3$. The heat flux profiles calculated using these $T_i$ measurements combined with data from the Langmuir probes give a comparison to that measured by infra-red camera.

PACS: 52.55 Fa, 52.55 Rk, 52.70 Ds

**\*Corresponding author address: EURATOM/CCFE Fusion Association, Culham Science Centre, Abingdon, Oxon, OX14 3DB, UK**

**\*Corresponding author E-mail: Sarah.Elmore@ccfe.ac.uk**

**Presenting author: Sarah Elmore**

**Presenting author e-mail: Sarah.Elmore@ccfe.ac.uk**

## 1. Introduction

One of the main interests in tokamak edge physics research is the understanding and characterisation of energy and particle fluxes to the divertor and plasma facing components. The target ion energy distribution (as might be characterised through temperature) is of particular concern for future large scale devices such as ITER, since it will be the ion energies which determine the damage by physical sputtering to the divertor target and plasma facing components [1]. The electron temperature ($T_e$) at the target is routinely measured using flush mounted Langmuir probes, however, measurements of ion energies and temperatures are more difficult to make [2]. In the absence of ion temperature ($T_i$) data, it is often assumed that $T_i = T_e$ when power fluxes to the divertor target are calculated using Langmuir probes. In the Mega Amp Spherical Tokamak (MAST), using this assumption, power balance has been found in L-mode and it has been confirmed by initial target ion temperature measurements which show that $T_i = T_e$ [3]. However, in inter-ELM H-mode, Langmuir probe data do not give power balance if $T_i = T_e$ is assumed [4]. The ratio of $T_i/T_e$ is therefore of interest at the target in inter-ELM H-mode.

In this paper divertor target $T_i$ measurements, from a retarding field energy analyser (RFEA), are presented during inter-ELM H-mode discharges on MAST. In section 2, the experimental setup is described, and in section 3 the target inter-ELM H-mode $T_i$ measurements are compared to the $T_e$ measurements from Langmuir probes. The effect of $T_i/T_e$ on power balance by probe theory is discussed and heat flux profiles are compared to infra-red (IR) data.

## 2. Experimental Setup

Ion temperature measurements in the MAST scrape-off layer (SOL) at the target were performed using a novel divertor RFEA [3] based on the design of the MAST midplane RFEA [5] (similar RFEAs have been used in a number of other machines [2, 6-11] and ion temperatures have been measured at the divertor of the Large Helical Device (LHD) by ion sensitive probes [12]). The divertor RFEA is installed on the Divertor Science Facility [13] at the lower outer divertor of MAST. Figure 1 (b) shows the location of the divertor RFEA in a cross-section of the MAST strike point region. The analyser is equipped with one module, identical to those in the MAST midplane RFEA [3], facing the plasma flow to the target.

The RFEA consists of a negatively biased front slit plate, a discriminator grid swept to a high positive voltage (grid 1 in figure 1 (a)), a negatively biased electron repelling grid (grid 2 in figure 1 (a)) and a grounded collector plate. Ions with energies higher than $eV_{grid1}$, where e is the ion charge and $V_{grid1}$ is the discriminator voltage, can overcome the coulomb repulsion and contribute to the collector current measured, $I_{col}$. The $I_{col}$-$V_{grid1}$ characteristic is related to the ion energy distribution so that, assuming a Maxwellian distribution of ions, $T_i$ can be extracted from a fit to the decaying part of the $I_{col}$-$V_{grid1}$ characteristic where $V_{grid1}$ is larger than the sheath voltage $V_s$ [3];

$$I_{col} = I_0 \exp\left[ -\frac{Z_i}{T_i}\left( V_{grid1} - |V_s| \right) \right] + I_{off} \, , \qquad (1)$$

where $I_{off}$ accounts for any current offsets. For the discharges discussed here the slit plate and electron repelling grid were held constant at -100 V and -175 V respectively. The discriminator grid was swept to a maximum of 120 V in 0.5 ms.

Although the divertor RFEA is installed at a fixed location, R = 0.985 m, radial profiles of the ion temperature are possible due to the natural movement of the strike point across the MAST divertor target. Radial $T_e$ profiles at the target have been measured for all discharges by an array of Langmuir probes and averaged over the time period when RFEA data is taken to show a comparison to the $T_i$ data.

## 3. Target ion temperature measurements in H-mode

During inter-ELM H-mode on MAST the Langmuir probes at all four strike points have been used to calculate the power to the divertor using $P_{DIV} = \gamma J_{sat} T_e$, where $\gamma = 5 + 2T_i/T_e$ [14] and $J_{sat}$ is the saturation current density integrated over the power deposition area. This is compared with the power entering the SOL, calculated from $P_{SOL} = P_{in} - \dot{W} - P_{rad}$ (see figure 2), where $P_{in}$ is the sum of the ohmic and absorbed neutral beam power, $\dot{W}$ is the rate of change of the stored plasma energy and $P_{rad}$ is the radiated power measured by bolometry [15]. If $T_i = T_e$ is assumed when calculating the power at probes from $P_{DIV}$ then the total power arriving at the divertor is less than the power entering the SOL. As shown in figure 2, power balance could be achieved if $T_i/T_e = 2 - 4$ was used to calculate power at probes. In order to test this, ion temperature measurements have been made for a series of discharges.

The first discharges studied were in a double-null configuration with plasma current, $I_p = 900$ kA at two neutral beam heating powers, $P_{NBI} = 1.5$ MW and $P_{NBI} = 3.4$ MW [16]. The discharge with $P_{NBI} = 1.5$ MW has regular type III ELMs and the $T_i$ data are extracted from the inter-ELM periods of the discharge. The discharge with $P_{NBI} = 3.4$ MW has an extended ELM free period during which the data were collected.

Figure 3 shows $T_i$ from the RFEA and $T_e$ from Langmuir probes as a function of distance from the last closed flux surface (LCFS) at the target for both discharges. The $P_{NBI} = 3.4$ MW discharge shows a higher $T_i$ range of $10 - 20$ eV compared to the $P_{NBI} = 1.5$ MW discharge which has $T_i = 8 - 12$ eV. However, both discharges have similar $T_e$ across the radial profile in the range $T_e = 4 - 8$ eV. This results in a $T_i/T_e \sim 2$, dropping slightly to $T_i/T_e \sim 1.4$ at $\Delta R_{LCFS}^{tgt} > 0.9$ m for the $P_{NBI} = 1.5$ MW discharge and $T_i/T_e \sim 2 - 3$ for the $P_{NBI} = 3.4$ MW. The $T_i$ measurements are restricted to $4 - 10$ cm from the LCFS because, whilst the plasma was in H-mode, the strike point only moved this distance relative to the divertor RFEA. This means the trend of increasing $T_i$, and therefore $T_i/T_e$, for the $P_{NBI}$ = 3.4 MW discharge cannot be confirmed at the LCFS.

Figure 4 shows the heat flux to the lower outer divertor target for the $P_{NBI}$ = 3.4 MW discharge during the extended inter-ELM period from IR camera data and the heat flux calculated from Langmuir probes, assuming $T_i = T_e$. Since the Langmuir probe heat flux does not match the IR data at the strike point when $T_i = T_e$ is assumed it is likely $T_i/T_e > 1$ for this region of the radial profile. In order to compare the heat flux calculated using the measured $T_i$ data, a fit to the data in the region 4 - 10 cm from the LCFS at the target has been used to produce a heat flux in the divertor SOL; this is also shown in figure 4. Near the strike point, the LP calculation gives an underestimate of q, in this region; although it is an overestimate, the heat flux using the RFEA data, considering error bars, overlaps with the IR data. Therefore $T_i/T_e > 1$ would give an improved match to the IR data at the strike point.

One possible hypothesis could be that the larger difference in the ion and electron temperatures, shown in figure 3, for the $P_{NBI}$ = 3.4 MW discharge compared to the $P_{NBI}$ = 1.5 MW discharge can be explained by the difference in collisionality. The higher beam power discharge has a collisionality (calculated at the upstream SOL using $v_{SOL,LCFS}^* = 10^{-16} n_e L / T_e^2$ where L is the connection length in the SOL [14]) of $v_{SOL,LCFS}^* \sim 12.2$, whereas the lower beam power has a higher collisionality of $v_{SOL,LCFS}^* \sim 15.3$. Therefore we would expect that for the higher collisionality, ions and electrons are more likely to equilibrate along the SOL towards the target, compared to the lower collisionality. An alternative hypothesis is that the ratio of $T_i/T_e$ at the target is dependent on $T_i/T_e$ or $\lambda_{Te}$ and $\lambda_{Ti}$ at the midplane, however experimental measurements at the midplane for both of these discharges are not yet available.

A similar shaped double-null discharge with a lower plasma current, $I_p$ = 600 kA, with beam heating, $P_{NBI}$ = 3.4 MW, and regular type I ELMs has been studied [16]. The RFEA data has been extracted from inter-ELM periods in the discharge. Figure 5 shows the radial profile of $T_i$ from RFEA and $T_e$ from LP at the target. The radial range for the $T_i$ data is 4 – 15 cm from the LCFS. Unfortunately data closer to the LCFS is not available because the slit plate power supply fails at this proximity to the LCFS due to the high slit plate currents during type I ELMs. When there is no voltage applied to the slit plate, the RFEA probe would be space-charge limited and therefore this data is rejected [3]. $T_i$ ranges from 9 eV to 20 eV up to 4 cm from LCFS at the target with $T_e$ = 5 - 10 eV. In this lower current, $I_p$ = 600 kA, discharge $T_i$ is lower than in the $I_p$ = 900 kA discharge, however $T_e$ is comparatively higher which results in a ratio of $T_i/T_e$ = 1.5 – 2.5 across the radial profile, increasing towards the LCFS. The difference in the $T_i/T_e$ ratio for these two discharge types may again be related to the upstream SOL collisionality. The lower current, $I_p$ = 600 kA, discharge has a collisionality, $v^*_{SOL,LCFS}$ ~ 14.1, which is higher than in the higher current discharge ($v^*_{SOL,LCFS}$ ~ 12.2). This suggests that $T_i/T_e$ at the target decreases with increasing collisionality as may be expected.

Finally a lower single null discharge with $I_p$ = 600 kA, $P_{NBI}$ = 3.4 MW and regular type I ELMs has been studied [16]. Figure 6 shows the temperature profile of $T_i$ from RFEA and $T_e$ from LP as a function of distance from the LCFS at the target. Ion temperatures are in the range 7 – 14 eV over the range 2 - 11 cm from LCFS. The electron temperature is of a similar range to other inter-ELM H-mode discharges presented in this paper, with $T_e$ = 4 – 9 eV. This gives $T_i/T_e$ = 1 – 2, which is the lowest ratio between the ion and electron temperature in the discharges discussed, inspite of the fact that the upstream collisionality at the LCFS, $v^*_{SOL,LCFS}$ ~ 8.2.

Although $T_i/T_e$ at the target was expected to be influenced by the collisionality, the dependence in this series of discharges were not clear and further investigation is necessary.

## 4. Summary

Measurements of the ion temperature at the divertor target in the MAST SOL were made in inter-ELM H-mode in a range of plasma discharges by RFEA. $T_i$ at the target varied over the plasma discharges studied however $T_e$ measured by Langmuir probes did not appear to alter greatly. The ratio of $T_i/T_e$ at 5 cm from the LCFS was found to range from $T_i/T_e$ = 3 in a lower collisionality, $v^*_{SOL,LCFS} \sim 12.2$, double-null discharge (DND) with $I_p$ = 900 kA and $P_{NBI}$ = 3.4 MW to $T_i/T_e$ = 2 in a higher collisionality, $v^*_{SOL,LCFS} \sim 15.3$, DND with $I_p$ = 900 kA and $P_{NBI}$ = 1.5 MW, showing an apparent scaling with collisionality. The lower single-null discharge with $I_p$ = 600 kA and $P_{NBI}$ = 3.4 MW had the lowest $T_i/T_e \sim 1.3$ at 5 cm from the LCFS and $T_i \sim T_e$ in the far SOL. Using the $T_i$ information measured in one particular MAST discharge condition, combined with the Langmuir probe data, allows heat fluxes to be calculated that are in closer agreement with those obtained from the IR camera at the strike point. This suggests that using $T_i = T_e$ does not allow power balance in inter-ELM H-mode but that higher $T_i/T_e$ ratios as measured by the divertor RFEA are more realistic.

## Acknowledgements

This work, part-funded by the European Communities under the contract of Association between EURATOM and CCFE, was carried out within the framework of the European Fusion Development Agreement. The views and opinions expressed herein do not necessarily reflect those of the European Commission. This work was also part-funded by the RCUK Energy Programme under grant EP/I501045.

**Figure captions**

Figure 1. a) Schematic of RFEA module showing function of components. b) Cross-section of MAST strike point, showing the location of the divertor RFEA at the lower outer target.

Figure 2. Inter-ELM H-mode energy balance showing the power into the SOL, $P_{SOL}$, for a number of discharges compared to the power arriving to the target Langmuir probes, power at probes, at all four strike points in MAST. The data is shown when assuming $T_i = T_e$ (black circles), $T_i = 2T_e$ (red squares) and $T_i = 4T_e$ (blue triangles) when calculating the power at probes. The line for energy balance can be seen to fall between $T_i = 2T_e$ and $T_i = 4T_e$, suggesting that the ion temperature is higher than the electron temperature by a factor of $2 - 4$.

Figure 3. $I_p = 900$ kA inter-ELM H-mode double-null discharge with $T_i$ measurements from the RFEA for $P_{NBI} = 1.5$ MW (solid black circles) and for $P_{NBI} = 3.4$ MW (solid red squares) and $T_e$ from Langmuir probes for $P_{NBI} = 1.5$ MW (hollow black circles) and $P_{NBI} = 3.5$ MW (hollow red squares), as a function of distance from the LCFS at the target. $T_e$ at both neutral beam injection (NBI) heating powers are relatively constant; however, $T_i$ at higher NBI heating power increases with proximity to the LCFS unlike the lower beam power case.

Figure 4. Heat flux profile at the target for $I_p = 900$ kA $P_{NBI} = 3.4$ MW inter-ELM H-mode discharge measured by infra-red camera (hollow red circles) and Langmuir probe measurements (solid black circles) assuming $T_i = T_e$. Also shown is the heat flux calculated from Langmuir probe measurements when using a fit to the measured RFEA data for $T_i$ in the SOL with LP measurements (hollow blue squares). It can be seen that when the Langmuir probe data is used to calculate heat flux with the correct value for $T_i/T_e$ the measurements increase to above the IR data.

Figure 5. $I_p$ = 600 kA $P_{NBI}$ = 3.4 MW inter-ELM H-mode double-null discharge with $T_i$ measurements from RFEA (solid black circles) and $T_e$ from Langmuir probes (hollow red squares), as a function of distance from the LCFS at the target. $T_i > T_e$ can be seen generally across the radial profile with $T_i/T_{e,max}$ ~ 2.5.

Figure 6. $I_p$ = 600 kA $P_{NBI}$ = 3.4 MW inter-ELM H-mode lower single-null discharge with $T_i$ measurements from RFEA (solid black circles) and $T_e$ from Langmuir probes (hollow red squares), as a function of distance from the LCFS at the target. $T_i \geq T_e$ can be seen generally across the radial profile with $T_i/T_{e,max}$ ~ 2.

**Figures**

**Figure 1**

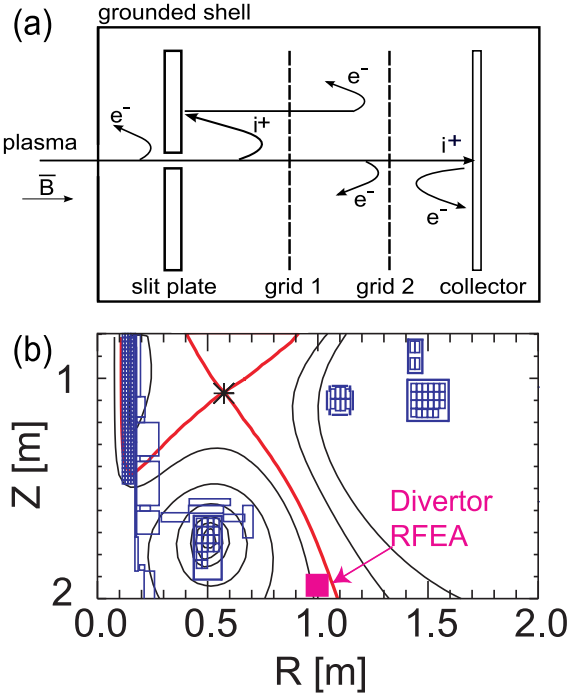

**Figure 2**

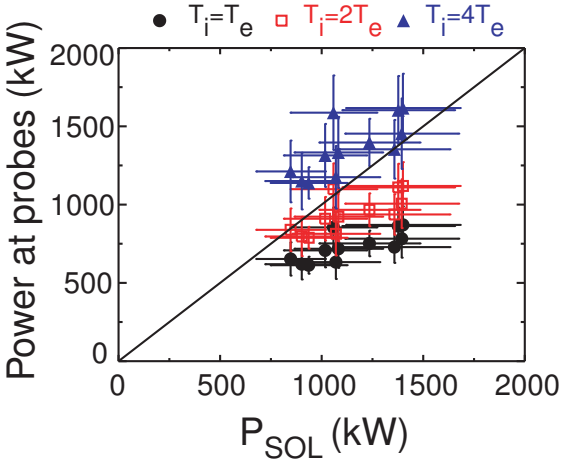

**Figure 3**

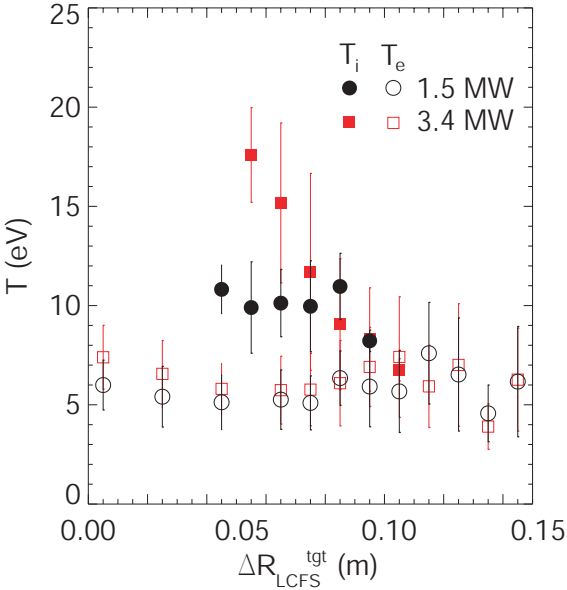

**Figure 4**

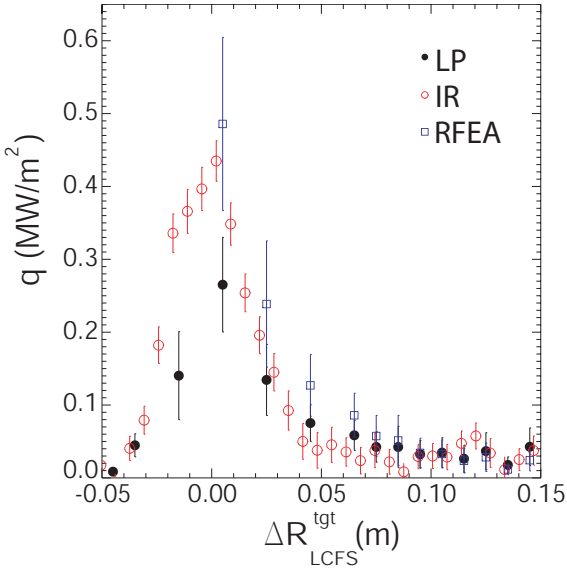

**Figure 5**

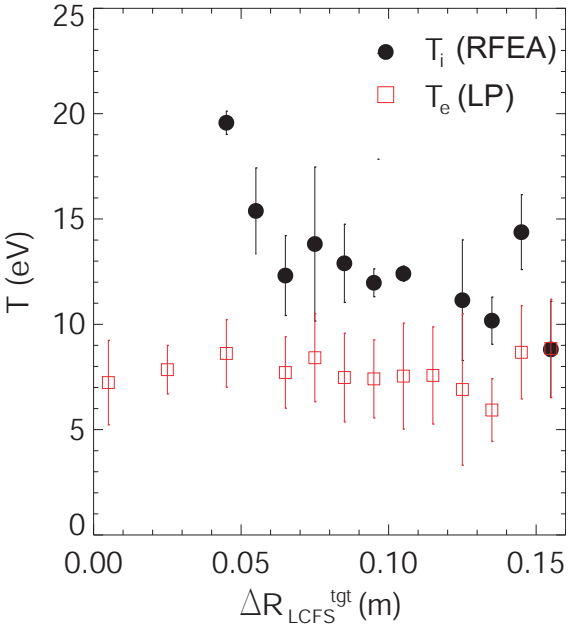

**Figure 6**

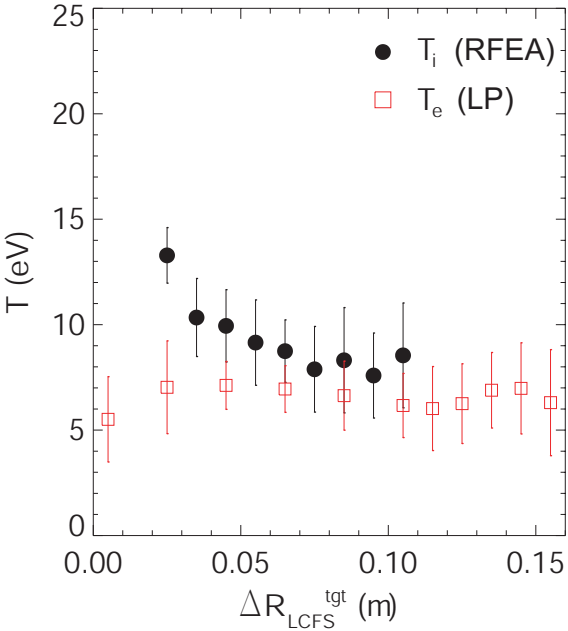